\documentclass[prl,twocolumn,showpacs,floatfix,amsmath,amssymb,superscriptaddress]{revtex4-2}
\usepackage{amsfonts}
\usepackage{stmaryrd}
\usepackage{bbm}
\usepackage{mathrsfs}
\usepackage{tipa}
\usepackage{amssymb}
\usepackage{txfonts}
\usepackage{graphicx}
\usepackage{dcolumn}
\usepackage{epstopdf}
\usepackage[colorlinks,linkcolor=blue,urlcolor=blue,citecolor=blue]{hyperref}
\usepackage{multirow}
\usepackage{subfigure}
\usepackage{url}
\usepackage{upgreek}
\usepackage[utf8]{inputenc}
\usepackage[english]{babel}
\usepackage{soul}
\usepackage[normalem]{ulem}

\begin{document}

\newcommand*{\cm}{cm$^{-1}$\,}
\newcommand*{\Tc}{T$_c$\,}
\newcommand{\ve}[1]{\boldsymbol{#1}}
\newcommand{\FA}[1]{{\color{magenta} #1}}

\title{Magnetization process of a quasi-two-dimensional quantum magnet:\\ Two-step symmetry restoration and dimensional reduction}

\author{Anneke~Reinold}
\affiliation{Department of Physics, TU Dortmund University, 44227 Dortmund, Germany}

\author{Lucas Berger}
\affiliation{Institute of Physics II, University of Cologne, 50937 Cologne, Germany}

\author{Marcin Raczkowski}
\affiliation{Institut für Theoretische Physik und Astrophysik, Universität Würzburg, 97074 Würzburg, Germany}

\author{Zhiying~Zhao}
\affiliation{State Key Laboratory of Structural Chemistry, Fujian Institute of Research on the Structure of Matter,
Chinese Academy of Sciences, Fuzhou, Fujian 350002, China}

\author{Yoshimitsu~Kohama}
\author{Masaki Gen}
\affiliation{Institute for Solid State Physics, University of Tokyo, Kashiwa, Chiba 277-8581, Japan}

\author{Denis~I.~Gorbunov}
\author{Yurii~Skourski}
\author{Sergei~Zherlitsyn}
\affiliation{Dresden High Magnetic Field Laboratory (HLD-EMFL), Helmholtz-Zentrum Dresden-Rossendorf, 01328 Dresden, Germany}

\author{Fakher F. Assaad}
\affiliation{Institut für Theoretische Physik und Astrophysik, Universität Würzburg, 97074 Würzburg, Germany}
\affiliation{Würzburg-Dresden Cluster of Excellence ct.qmat, Am Hubland, D-97074 Würzburg, Germany}

\author{Thomas Lorenz}
\affiliation{Institute of Physics II, University of Cologne, 50937 Cologne, Germany}

\author{Zhe Wang}
\affiliation{Department of Physics, TU Dortmund University, 44227 Dortmund, Germany}

\date{\today}

\begin{abstract}
We report on a comprehensive thermodynamic study of a quasi-two-dimensional (quasi-2D) quantum magnet Cu$_2$(OH)$_3$Br which in the 2D layer can be viewed as strongly coupled alternating antiferromagnetic and ferromagnetic chains.
In an applied magnetic field transverse to the ordered spins below $T_N=9.3$~K, a field-induced phase transition from the 3D ordered to a disordered phase occurs at $B_c=16.3$~T for the lowest temperature, which is featured by an onset of a one-half plateau-like magnetization. 
By performing quantum Monte Carlo simulations of the relevant 2D model, we find that the plateau-like magnetization corresponds to a partial symmetry restoration and the full polarization in the ferromagnetic chains. 
Our numerical simulations also show that the magnetization saturation occurs with full symmetry restoration at a much higher field of $B_s \simeq 95$~T, corresponding to a 1D quantum phase transition in the antiferromagnetic chains. 
We argue that the experimentally observed field-induced phase transition at $B_c$ follows from the partial symmetry restoration and the concomitant dimensional reduction.
\end{abstract}

\maketitle

Quantum matter in magnetic fields is a long standing active domain of research, which exhibits fascinating quantum phenomena such as quantum phase transitions \cite{Sachdev} and magnetization plateaus \cite{Chubukov1991,Starykh10,Ye17,Yamaguchi18,Bachus20}. 
The nature of the magnetization processes often reveals the underlying physics. 
For the two-dimensional (2D) Heisenberg models, the occurrence of a phase transition depends on the sign of the exchange coupling. For an antiferromagnetic coupling the transition to the fully polarized state falls in the universality class of magnonic Bose-Einstein condensation \cite{Batyev84,Giamarchi08,Matsumoto24}. Before the transition the magnetic field first induces a canted antiferromagnet, where the rotational  spin symmetry around  the magnetic field axis is  spontaneously broken, whereas in the fully polarized state this symmetry is restored. 
For the ferromagnetic coupling,  the order parameter - the total spin -  will directly align with the magnetic field  such  that no  transition at finite magnetic field strength occurs.

\begin{figure*}[t]
\centering
\includegraphics[width=1\linewidth]{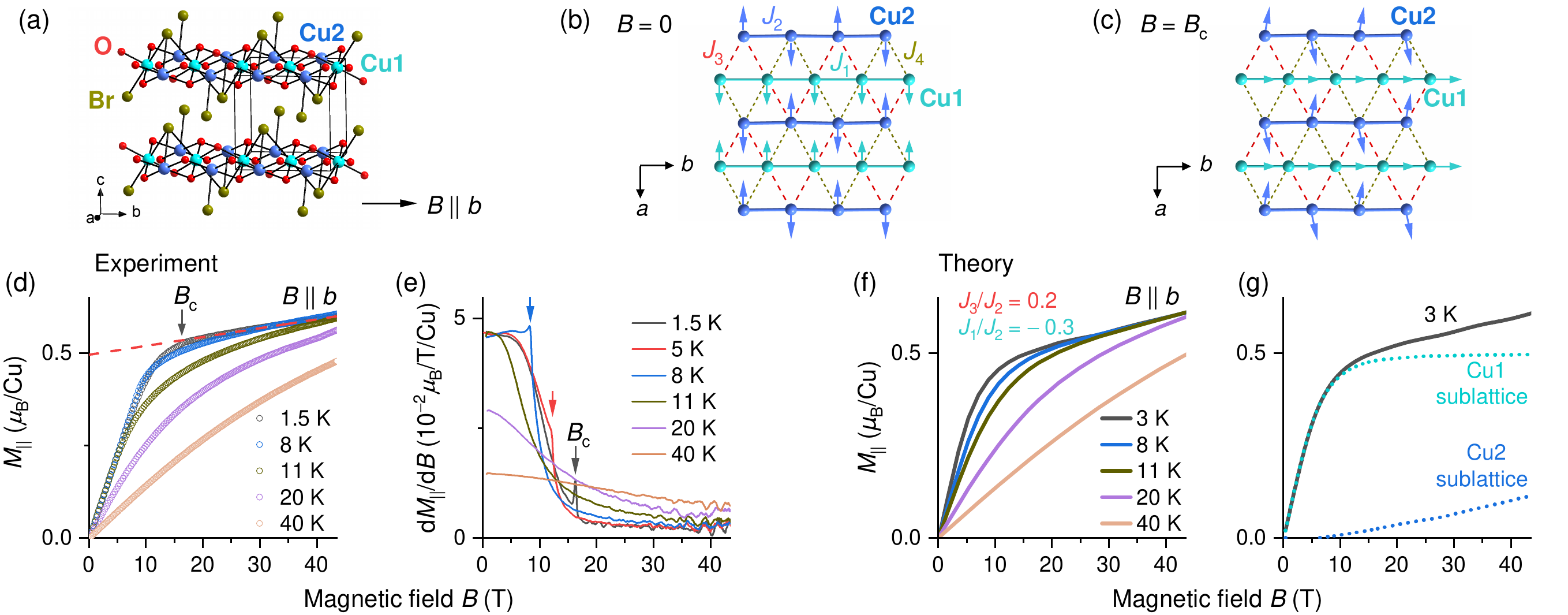}
\caption{
(a) Layered crystal structure. 
(b) Illustration of a long-range order at zero field, with the ordered Cu spins projected in the \textit{ab} layer (see text for more details).
(c) Illustration of a peculiar magnetic structure induced by an applied field $B=B_c$ along the \textit{b} axis: the Cu1 spins are fully polarized, while the Cu2 spins remain nearly intact.
(d) Magnetization and (e) its field derivative versus field at various temperatures.
Theoretically calculated magnetization (f) at various temperatures and (g) at 3~K with the sublattice contributions separately evaluated.
}
\label{fig:Struc}
\end{figure*}

The motivation of our experimental and theoretical study in this work is to investigate  the  magnetization process of a system that exhibits both antiferromagnetic and  ferromagnetic correlations.
A quasi-2D magnet is realized in a Botallackite compound Cu$_2$(OH)$_3$Br based on spin-1/2 Cu$^{2+}$ ions with space group $P2_1/m$ [see Fig.~\ref{fig:Struc}(a)] \cite{Aebi48,OswaldLudi61}.
The Cu ions have two different crystallographic local environments with the corresponding Cu1 and Cu2 spins forming chains directed along the crystallographic \textit{b} axis, and the two types of chains alternate  along the \textit{a} axis \cite{Aebi48,OswaldLudi61,ZhengKawae09,ZhaoHe19,ZhangKe20,XiaoTong22}. The spins along the Cu2 chain are coupled by a nearest-neighbor antiferromagnetic exchange $J_2$, while in the Cu1 chain by a smaller ferromagnetic $J_1$.
The Cu1 and Cu2 chains are coupled by interchain couplings $J_3$ and $J_4$, leading to an overall triangular arrangement of the Cu spins in the \textit{ab} plane [Fig.~\ref{fig:Struc}(b)].

If $J_3$,$J_4 \ll J_1$,$J_2$, we will have a well-understood weakly-coupled spin chain system, where the magnetic properties are governed by the intrachain $J_1$ and $J_2$ interactions.
However, in Cu$_2$(OH)$_3$Br while $J_4$ is negligibly small, the interchain coupling $J_3$ is not small, but comparable to $J_1$, i.e. $|J_3| \simeq |J_1| < |J_2|$ \cite{ZhangKe20}.
In addition, an even smaller inter-layer coupling stabilizes a long-range magnetic order at a quite low temperature $T_N = 9.3$~K \cite{ZhengKawae09,ZhaoHe19,ZhangKe20,XiaoTong22}, although the leading intra-layer interaction $J_2/k_B = 61.5$~K is much greater. 
Therefore, Cu$_2$(OH)$_3$Br is a very good realization of a quasi-2D magnet with both antiferromagnetic and ferromagnetic correlations.

By performing thermodynamic measurements, we find that an applied transverse magnetic field can suppress the long-range spin order. While the phase transition corresponds to a field-induced polarization of the Cu1 spins, the critical field $B_c$ is determined by the \textit{interchain} coupling $J_3$ (rather than the intrachain $J_1$) in competition with the applied field. 
By carrying out quantum Monte Carlo simulations of the underlying 2D model, we show that the field-induced phase transition corresponds to a crossover to an intermediate phase that is characterized by a 
\textit{partial} symmetry  restoration, in which the ground state is invariant under rotation of \textit{only} the Cu1 spins around the magnetic field axis. At a much higher magnetic field, the numerical simulations reveal a quantum phase transition to the fully polarized state, where the symmetry is completely restored and the corresponding critical exponents are of one dimensionality despite in a 2D magnet.

High-quality Cu$_2$(OH)$_3$Br single crystals with a typical size of $\sim 2 \times 5 \times 0.5\,$mm$^3$ were grown by using a conventional hydrothermal method \cite{Aebi48,OswaldLudi61,ZhengKawae09,ZhaoHe19} .
Magnetization measurements were performed down to 1.5~K and in static and pulsed magnetic fields up to 14 and 56~T, respectively.
Magnetocaloric-effect experiment was carried out in pulsed fields up to 49~T for temperatures down to 1.3~K.
For ultrasound experiment~\cite{Luethi05,Hauspurg24}, a cryogenic superconducting magnet for temperatures down to 2~K and fields up to 14~T was utilized, 
while thermal expansion and magnetostriction were measured with a capacitance dilatometer 
down to 0.3 K using a $^3$He insert in a 17 T magnet. The field and temperature dependence
of heat capacity was measured using the calorimeter of the Physical Property Measurement System.

\begin{figure*}[t]
\centering
\includegraphics[width=1\linewidth]{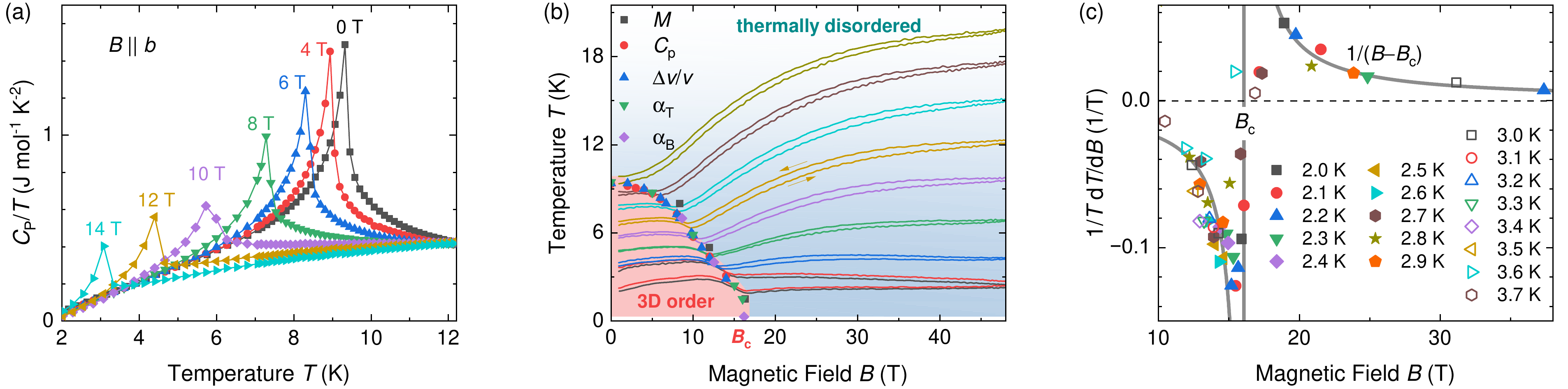}
\caption{
(a) Specific heat $C_p/T$ at various fields. 
(b) Magnetocaloric-effect measurements determine a phase boundary, consistent with the field-induced anomalies in magnetization $M$, specific heat $C_P$, sound velocity $\Delta v/v$, thermal-expansion coefficient $\alpha_T$, and magnetostriction coefficient $\alpha_B$.
(c) Derived magnetic Grüneisen parameter $\frac{1}{T}\frac{dT}{dB}$ exhibits a divergent behavior following $\propto 1/(B-B_c)$ with $B_c=16.1 \pm 0.8$~T. 
}
\label{fig:PD}
\end{figure*}

The long-range magnetic order below $T_N = 9.3$~K \cite{ZhengKawae09,ZhaoHe19,ZhangKe20,XiaoTong22} is characterized by the Cu2 spins being ordered along the $\pm\, a$ axis and the Cu1 spins nearly along the diagonals of the $ac$ planes [see Fig.~\ref{fig:Struc}(b) with the projection into the \textit{ab} plane] \cite{ZhangKe20}.
By applying a field transverse to the ordered spins at zero field, i.e. $B \parallel b$, we measure the longitudinal magnetization $M_\parallel \parallel b$ at various temperatures.
The low-temperature magnetization curves start with a steep increase, which
rapidly becomes gentle in the field range around 10~T [Fig.~\ref{fig:Struc}(d)]. 
At 1.5~K the $M_{\parallel}(B)$ curve exhibits a kink at $B_c=16.3\,$T, corresponding to a sharp peak in the differential susceptibility $dM_\parallel/dB$ [Fig.~\ref{fig:Struc}(e)].
Similar anomalies are visible in $dM_\parallel/dB$ at higher temperatures, which shift monotonically to lower fields [see arrows in Fig.~\ref{fig:Struc}(e)] with increasing $T$ and finally disappear above $T_N$. It is also interesting to note that the initial slope of $M_{\parallel}(B)$ remains essentially temperature independent up to $11\,$K that is already above $T_N$.
With further increasing temperature both, the initial slope and the overall curvature of the magnetization curves continuously decrease.

At high fields above about 35~T, all low-temperature magnetization curves approach a finite paramagnetic-like differential susceptibility of about $2.5\times 10^{-3}\,\mu_B$/T/Cu [Fig.~\ref{fig:Struc}(e)]. 
By extrapolating the high-field slope back to zero field, we obtain a value corresponding to $\mu_B/2$ per Cu, see dashed line in Fig.~\ref{fig:Struc}(d). 
Overall, the plateau-like magnetization in combination with the field-dependent suppression of the anomalies in the low-temperature differential susceptibility strongly indicates a field-induced phase transition at $B_c$. However, an intermediate plateau at $\mu_B/2$ per Cu is in clear contrast to the expected 1/3 plateau of the isotropic triangular-lattice antiferromagnet \cite{Starykh10} and also different from the saturation value of $1\,\mu_B$/Cu of polarized spin chains \cite{Pfeuty70,Breunig17}.

A suppression of the magnetically ordered phase by $B \parallel b$ is confirmed by measuring heat capacity $C_p$.
The peak position of the $C_p/T$ curve observed at $T_N=9.3$~K in zero field [Fig.~\ref{fig:PD}(a)], corresponding to the formation of the long-range order \cite{ZhengKawae09,ZhaoHe19,ZhangKe20,XiaoTong22}, 
shifts monotonically towards lower temperatures with increasing field.  
This clearly indicates a systematic suppression of the long-range order due to the competition of the Zeeman energy with the exchange interactions. Extrapolation of $T_N(B)$ suggests a vanishing of this energy scale at $B_c \simeq $ 16 T, indicating a field-induced quantum phase transition from a 3D ordered to a disordered phase.

We obtain further characteristics of the field-induced quantum phase transition by measuring magnetocaloric effect (MCE) in higher pulsed fields [Fig.~\ref{fig:PD}(b)].
Starting from the lowest temperature at zero field, the sample temperature increases slightly with field and then exhibits a minimum around 16~T, which is followed by a tiny increase towards highest field. In the corresponding down sweep of the magnetic field, the temperature is slightly elevated and exhibits a minimum at about 15~T.
With enhanced zero-field start temperature up to $T_N$, the field-dependent temperature curves $T(B)$ exhibit always a minimum. The minimum position shifts to lower field monotonically, which determines a phase boundary between the 3D long-range order and the field-induced 
disordered phase. 

Since the MCE experiment is performed under quasi-adiabatic conditions, the decrease (or increase) of the sample temperature indicates a heat flow from (or to) the lattice degrees of freedom to (or from) the magnetic degrees of freedom.
Above $T_N$ in the disordered phase [Fig.~\ref{fig:PD}(b)], the applied field can polarize the spins, reducing the entropy of the magnetic degrees of freedom and leading to the observed monotonic increase of the lattice temperature.
In contrast, the minimum exhibited by the lower-temperature $T(B)$ curves reflects an entropy accumulation in the magnetic subsystem, corresponding to enhanced thermal and/or quantum spin fluctuations. 
Because thermal fluctuations freeze with decreasing temperature, on approaching a quantum phase transition the finite-temperature behavior is expected to be finally dominated by the enhancing quantum fluctuations.
Experimentally it can be challenging to separate quantum completely from thermal fluctuations. 
Nonetheless, for magnetic systems the magnetic Grüneisen parameter $\Gamma_B(B,T)=\frac{1}{T}\frac{dT}{dB}$ turns out to exhibit a sign change with characteristic divergences upon approaching the quantum phase transition either as a function of temperature or magnetic field~\cite{ZhuSi03,Garst2005,Lorenz2008,Breunig17,WangLoidl18}. Based on the MCE data in Fig.~\ref{fig:PD}(b) we derive $\Gamma_B(B,T)$ and display the data in Fig.~\ref{fig:PD}(c) for the lowest temperatures. 
On approaching the phase boundary from higher field, the Grüneisen parameter exhibits a divergent behavior
$\propto 1/(B-B_c)$
with $B_c = 16.1 \pm 0.8 $~T in good agreement with the critical field determined from the magnetization data, and with further decreasing field $\Gamma_B$ shows the expected sign change, which provides the thermodynamic evidence for a field-induced quantum phase transition.

\begin{figure}[t]
\centering
\includegraphics[width=1\linewidth]{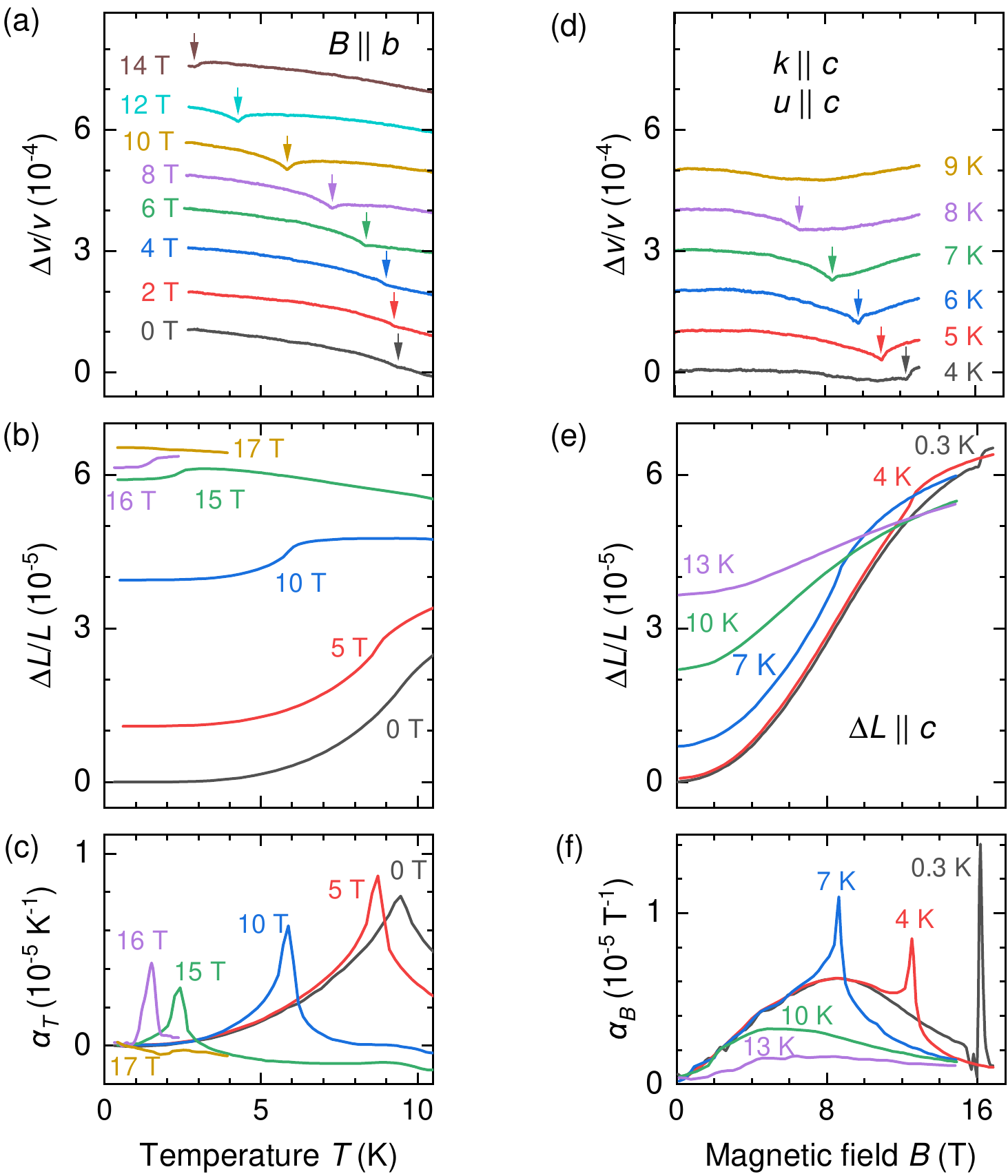}  
\caption{
(a) Ultrasound velocity $\Delta v(T)/v $ at 77~MHz, 
(b) thermal expansion $\Delta L(T)/L $ and 
(c) thermal expansion coefficient $\alpha (T)$ measured as a function of temperature at various fields. 
(d) Ultrasound velocity $\Delta v(B)/v $, (e) magnetostriction $\Delta L(B)/L $ and (f) magnetostriction coefficient $\alpha (B)$ versus field at various temperatures.
The curves in (a) and (d) are shifted vertically for clarity.
}
\label{fig:inelastic}
\end{figure}

Furthermore, we investigate the magneto-lattice responses by measuring ultrasound velocity change $\Delta v/v$ and the uniaxial thermal expansion $\Delta L/L$ , both along the crystallographic $c$ axis. 
As shown in Fig.~\ref{fig:inelastic}(a),  $\Delta v/v$ for a longitudinal polarization $u \parallel k \parallel c$ generally increases, i.e. the lattice hardens, with decreasing temperature. 
The magnetic ordering is indicated by a small dip at $T_N(B)$ as marked by the arrows, which continuously shifts to lower temperature with increasing field, and the corresponding dips are seen also in the field dependent data of $\Delta v/v$ [Fig.~\ref{fig:inelastic}(d)]. 

Figure~\ref{fig:inelastic}(b) shows that the long-range magnetic order causes a spontaneous contraction $\Delta L(T)/L$ of the $c$ axis, which corresponds to the interlayer distance, and $T_N$ is signalled by a sharp peak in the thermal expansion coefficient $\alpha_T \equiv \frac{1}{L}\frac{d\Delta L}{dT}$ [Fig.~\ref{fig:inelastic}(c)].
The continuous shift of $T_N(B)$ is again clearly visible, and the spontaneous contraction strongly reduces for $ B \geq 10\,$T. In contrast, the peaks in $\alpha_T$ hardly change up to 16~T, but completely vanish for $B = 17\,$T~$>B_c$. 
At 17~T, the sign of $\alpha_T$ becomes positive and the $\alpha_T(T)$ curve is nearly featureless down to the lowest temperature, which again reflects the expected behavior upon crossing a quantum phase transition~\cite{Garst2005,Breunig17}. 

The field-dependent $\Delta L(B)/L$ data and the corresponding magnetostriction coefficient $\alpha_B \equiv \frac{1}{L}\frac{d\Delta L}{dB}$ are displayed in  Figs.~\ref{fig:inelastic}(e) and \ref{fig:inelastic}(f), respectively. For $T>T_N$, the $\alpha_B(B)$ curves exhibit broad maxima, which increase with decreasing $T$, whereas all curves for $T<T_N$ seem to follow a universal curve up to the temperature dependent critical field $B_c(T)$ where the transition to the disordered phase causes a sharp peak of  $\alpha_B$. Most interestingly, the extremely sharp peak at the lowest $T=0.3\,$K appears right at $B_c=16.1\,$T, and the peak shape already resembles a weakly first-order transition with a jump-like discontinuous length change $\Delta L(B)/L$. Such a change from continuous to discontinuous transitions could arise from the finite magnetoelastic coupling as is discussed in the context of quantum phase transitions, e.g.\ in \cite{grigera2001,Thede2014,Zacharias2015,Belitz2017,Stapmanns2018}.

To address the microscopic magnetization process corresponding to the quantum phase transition, we consider a 2D spin-1/2 Heisenberg triangular lattice model
$\hat{H} = \sum_{i,j}J_{i,j}\ve{\hat{S}}_i\cdot\ve{\hat{S}}_j - g \mu_B B \sum_{i} S_{i}^{x}$ with the spin operators $\ve{\hat{S}}_i$ and $\ve{\hat{S}}_j$ denoting the pairs of Cu spins connected by the dominant exchange interactions $J_2$, $J_3$, or $J_1$  \cite{ZhangKe20}
[see Fig.~\ref{fig:Struc}(b)] and the Zeeman term with a Landé \textit{g}-factor and $B \parallel x \parallel b$ . 
With the much smaller $J_4$ omitted \cite{ZhangKe20}, which eliminates the negative sign problem,
we simulate the above Hamiltonian using the ALF (Algorithms for Lattice Fermions)  implementation of the  finite temperature auxiliary field quantum Monte Carlo method~\cite{ALF_v2,Blankenbecler81,Assaad08_rev}  and for
a $12\times 6$ system size with a unit cell comprising two Cu1 and two Cu2 spins. 
As for the Trotter discretization, we have used an imaginary time step $\Delta \tau J_2 \in [0.02,0.1]$ depending on the inverse temperature $\beta J_2 \in [1.54,20]$.
For the experimental parameters of Cu$_2$(OH)$_3$Br, i.e. $J_1/J_2=-0.3$, $J_3/J_2=0.2$, $J_2=5.3$~meV  \citep{ZhangKe20}, and $g=2.24$, the obtained magnetization curves  
well reproduce the experimental results at different temperatures [Fig.~\ref{fig:Struc}(d)(f)].
By evaluating the magnetization of the Cu1 and the Cu2 sublattices separately [Fig.~\ref{fig:Struc}(g)], we can clarify the magnetization process of the system:
The occurrence of the 1/2-plateau-like feature corresponds to a complete polarization in the Cu1 chains, whereas the Cu2 spins are only slightly polarized even up to 40~T [Fig.~\ref{fig:Struc}(c)], indicating a much higher saturation field.

\begin{figure}[t]
	\centering
	\includegraphics[width=1\linewidth]{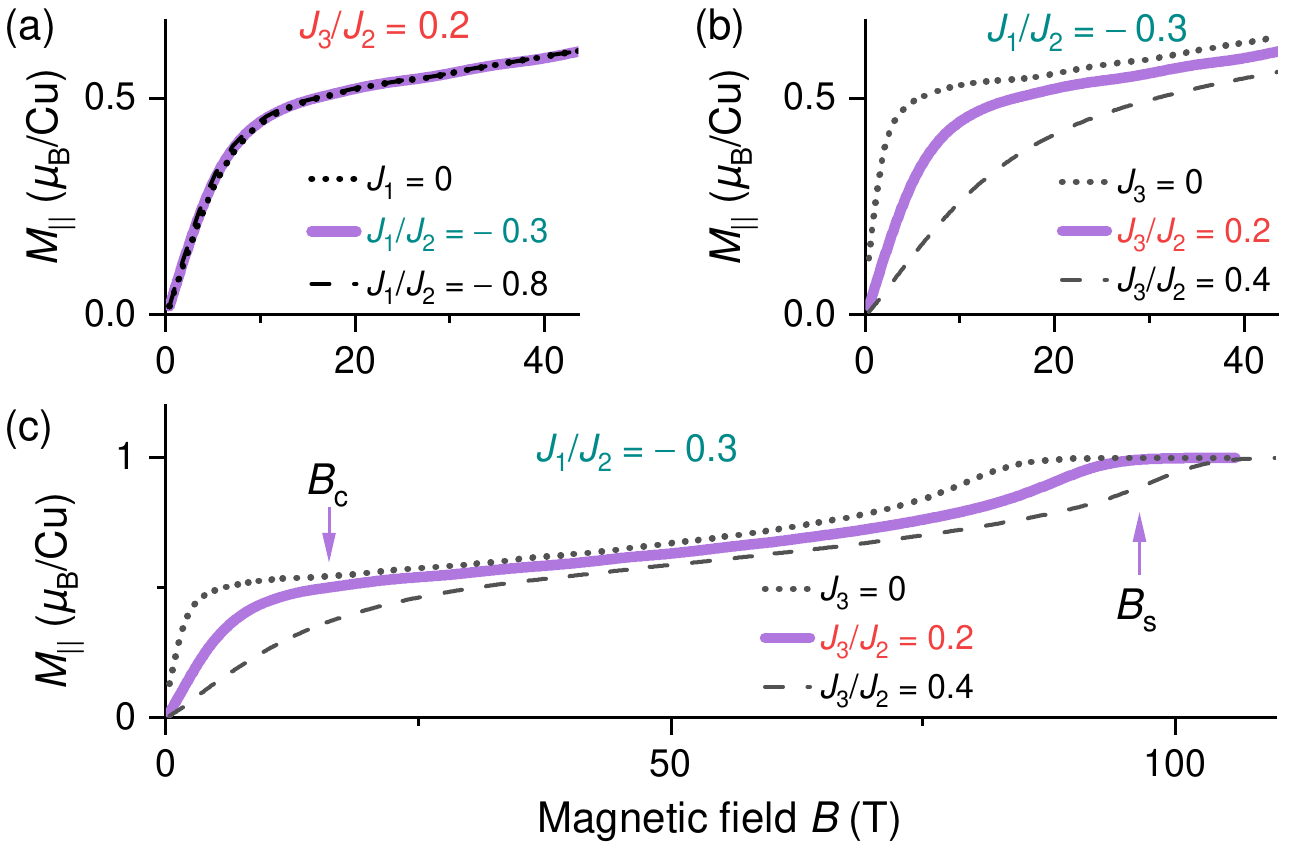}
	\caption{Quantum Monte Carlo simulation at 3~K.
		(a) For $J_3/J_2= 0.2$, the magnetization is essentially independent on $J_1$.
		(b) For $J_1/J_2 = -0.3$, the field-dependent magnetization is sensitive to the change of $J_3$.
			 (c) For the experimental parameters $J_3/J_2= 0.2$ and $J_1/J_2 = -0.3$, the simulated magnetization saturates at $B_s \simeq 95$~T.
	}
	\label{fig:theory}
\end{figure}

Such a magnetization process intuitively points to a magnetic structure of weakly coupled ferromagnetic and antiferromagnetic chains. 
However, this intuitive understanding is refuted by our theoretical simulations,  which reveal that the magnetization process is essentially independent from the ferromagnetic intra-chain coupling $J_1$ between the Cu1 spins. 
For the experimental value of $J_3/J_2=0.2$, the magnetization curves corresponding to $J_1/J_2 = 0$, $-0.3$ or $-0.8$ remain practically unchanged [Fig.~\ref{fig:theory}(a)]. In contrast, a variation of $J_3$ from $0.2J_2$ strongly modifies the magnetization [Fig.~\ref{fig:theory}(b)].

The best fit to the experimental results requires $J_3/J_2=0.2$  thus placing the model, from the point of view of coupling constants, well in the 2D regime. 
In a finite magnetic field $B \parallel x$, the SU(2) spin symmetry is broken, and one can decompose a spin into a longitudinal $\ve{S}_{\ve{i}}^{\parallel} = (\ve{e}_x \cdot \ve{S}_{\ve{i}}) \ve{e}_x $  and a transverse, $\ve{S}_{\ve{i}}^{\perp} = \ve{S}_{\ve{i}} - \ve{S}_{\ve{i}}^{\parallel}$ component. 
Whereas $\ve{S}_{\ve{i}}^{\parallel}$ is pinned by the field, the $\ve{S}_{\ve{i}}^{\perp}$ component fluctuates.
Upon entering the plateau-like state with fully polarized Cu1 spins, the U(1) symmetry along the field axis is partially restored: Since $\ve{S}_{\ve{i}}^{\perp}$ of the Cu1 spins vanishes, it is invariant under a global rotation of \textit{only} the Cu1 spins around the applied field. This provides a geometric route to decouple the spin chains and leads to a dimensional reduction of the transverse fluctuations.
This understanding is supported by our numerical simulation of the magnetization curve towards saturation, which is reached at $B_s \simeq 95$~T for the experimental parameters [see Fig.~\ref{fig:theory}(c)]
Approaching $B_s$ from below, the key signature of one dimensionality \cite{Breunig17}, i.e. a square root singularity $\propto 1-\sqrt{B_s-B}$, is observed numerically in this 2D system, in contrast to an expected linear field dependence for 2D fluctuations. 
Since our simulations are carried out at finite temperature, we observe a thermal rounding of this singular behavior, which is also seen in the decoupled case $J_3 =0$.
The effective magnetic field perceived by the Cu2 spins is given by $ g \mu_B B - 2SJ_3$, which explains the $J_3$ dependence of $B_s$ [Fig.~\ref{fig:theory}(c)].

While the 2D Heisenberg model does not exhibit a long-range order, the experimentally observed phase transition at $T_N$ is due to the weak inter-layer couplings.
The field-induced reduction of $T_N$ can be understood in terms of the decreasing transverse magnetization.
The experimentally observed vanishing of $T_N$ at $B_c \simeq 16$~T indicates a field-induced quantum phase transition, which roots in the enhanced quantum fluctuations due to the field-induced crossover from the quasi-2D to the 1D-like regime.
Since the experimental temperature down to 0.3~K is much lower than $J_2/k_B=61.5$~K, the field-induced disordered phase above $B_c$ is definitely governed by quantum fluctuations, whose spin dynamics is very interesting for further investigation.

To conclude, by performing comprehensive thermodynamic measurements of the quasi-two-dimensional quantum magnet  Cu$_2$(OH)$_3$Br as a function of magnetic field, and carrying out quantum Monte Carlo simulations of the underlying 2D model, we identify a rich set of signatures for a field-induced quantum phase transition, including a one-half plateau-like magnetization, suppression of Néel temperature, and field-induced dimensional reduction. 
In contrast to the 1/3-magnetization plateau for an isotropic triangular lattice or other geometrically frustrated antiferromagnets (e.g. \cite{Zhitomirsky02,Ono2004,Kamiya2018,ZhaoGegenwart20,Yamamoto2021,
Sheng22,Shangguan2023,Jeon2024,suetsugu2023emergent}),
the observed 1/2-plateau-like feature results from partial U(1) symmetry restoration and concomitant dimensional reduction.
Our work provides a novel route to study the physics of emergent 1D fluctuations in a (quasi-)2D system.

\begin{acknowledgments}
We thank H. O. Jeschke, X. Ke, C. Kollath, and F. Lisandrini for stimulating discussions, and acknowledge support by the European Research Council (ERC) under the Horizon 2020 research and innovation programme, Grant Agreement No. 950560 (DynaQuanta), and from the Deutsche Forschungsgemeinschaft (DFG, German Research Foundation) under Projects No. 277146847 - CRC 1238 (sub-projects No. A02, B01, B05) and No. 247310070 - SFB 1143, as well as the support of the HLD at HZDR, member of the European Magnetic Field Laboratory (EMFL).
M.R. is funded by the Deutsche Forschungsgemeinschaft (DFG, German Research Foundation), Project No. 332790403.
F.F.A. acknowledges financial support from the W\"urzburg-Dresden Cluster of Excellence on Complexity and
Topology in Quantum Matter ct.qmat (EXC 2147, Project No. 390858490).
M.R. and F.F.A. gratefully acknowledge the Gauss Centre for Supercomputing e.V. (www.gauss-centre.eu) for funding
this project by providing computing time on the GCS Supercomputer SuperMUC-NG at Leibniz Supercomputing Centre (www.lrz.de).
Z.Z. is financed by the National Natural Science Foundation of China (Grant No. 52072368)
\end{acknowledgments}

\bibliographystyle{apsrev4-2}
\bibliography{COHB_bib}

\end{document}